\documentclass[proceedings]{JHEP3} 

\usepackage{epsfig}                   
\usepackage{amsmath}
\usepackage{amssymb}

\newcommand{\AddrAHEP}{%
  AHEP Group, Instituto de F\'{\i}sica Corpuscular --
  C.S.I.C./Universitat de Val{\`e}ncia \\
  Edificio Institutos de Paterna, Apt 22085, E--46071 Valencia, Spain}

\PrHEP{PRHEP-AHEP2003/072}
\conference{International Workshop on Astroparticle and High Energy Physics}

\title{Efficient solar anti-neutrino production in random magnetic fields}

\author{O. G. Miranda${}^1$, T. I. Rashba${}^{2,3}$,
        \speaker{A. I. Rez}${}^{2,3}$ and J.~W.~F. Valle${}^2$\\

{${}^1$Departamento de F\'{\i}sica, Centro de Investigaci{\'o}n y de
  Estudios Avanzados, Apdo.~Postal 14-740 07000 Mexico, DF, Mexico}\\

{${}^2$\AddrAHEP}\\

{${}^3$Institute of Terrestrial Magnetism,
    Ionosphere and Radio Wave Propagation of the Russian Academy of Sciences,
    142190, Troitsk, Moscow region, Russia}\\

E-mail: \email{rez@izmiran.rssi.ru}}

\abstract{%
  We have shown that the electron anti-neutrino appearance in the
  framework of the spin flavor conversion mechanism is much more
  efficient in the case of neutrino propagation through random than
  regular magnetic field. This result leads to much stronger limits on
  the product of the neutrino transition magnetic moment and the solar
  magnetic field based on the recent KamLAND data. We argue that the
  existence of the random magnetic fields in the solar convective zone
  is a natural sequence of the convective zone magnetic field
  evolution.}

\begin{document}

\section{Introduction}
\label{sec:introduction}

Recently the KamLAND experiment has announced that the electron
anti-neutrino component in the solar flux is less than
$2.8\times10^{-2}\%$ of the solar boron flux at the $90\%$
C.L.~\cite{Eguchi:2003gg}, a bound about 30 times more stringent than
the latest Super-Kamiokande limit~\cite{Gando:2002ub}.

The presence of electron anti-neutrinos in the solar flux may indicate
the existence of spin-flavor precession (SFP) induced by non-vanishing
neutrino transition magnetic
moments~\cite{Schechter:1981hw,akhmedov:1988uk} interacting with solar
magnetic fields or, alternatively, neutrino decays in models with
spontaneous violation of lepton
number~\cite{schechter:1982cv,gelmini:1984ea,beacom:2002cb}.  Here we
discuss the case of anti-neutrinos produced by SFP conversions.

The first KamLAND evidence of reactor anti-neutrino
disappearance~\cite{Eguchi:2002dm} had already excluded SFP scenarios
as solutions to the solar neutrino problem~\cite{Barranco:2002te}.
Together with the latest KamLAND limit on the solar
electron anti-neutrino flux~\cite{Eguchi:2003gg} and including the recent SNO
salt results~\cite{Ahmed:2003kj} the robustness of the
LMA MSW solution to the solar neutrino problem (SNP)%
\footnote{For the recent analysis of the solar neutrino data after the
  SNO salt results~\cite{Ahmed:2003kj} in the simplest three-neutrino
  LMA MSW oscillation picture but neglecting neutrino magnetic moments
  effects see, e.g.,~\cite{Maltoni:2003da,maltoni:2002aw}.}  against
the SFP mechanism was confirmed~\cite{Miranda:2003yh}.  However, a
neutrino magnetic moment could still play a notable role and lead to
sub-leading, but potentially observable, effects.

To analyze the SFP conversion several solar magnetic field models were
considered previously, characterized by different assumptions
pertaining to their magnitude, location and typical scales, regular or
random
nature~\cite{Bykov:1998gv,Kutvitskii,Guzzo:1998sb,akhmedov:2002mf,Friedland:2002pg}
coming from our lack of knowledge about solar magnetic fields.
Usually the solar magnetic fields are supposed to reside within the
solar convective zone~\cite{Bykov:1998gv,Kutvitskii,Guzzo:1998sb} in
agreement with dynamo mechanism. Sometimes they are considered to be
located in the solar core or in the radiative
zone~\cite{akhmedov:2002mf,Friedland:2002pg}. Although allowed, these
are not physically as persuasive as the former ones.

In what follows we adopt the conservative point of view, assuming a
convective zone magnetic field model and exploiting the fact that in
accordance with the present-day understanding of solar magnetic field
evolution, the large-scale magnetic field in the solar convective zone
is followed by a small-scale random component, the strength of which
is comparable to or even larger than that of the regular one.

The main issue we advocate here is that within the SFP scenario, solar
random magnetic fields can generally result in a sizable gain in
electron anti-neutrino yield, up to one-two orders of magnitude as
compared to regular fields of the same (in the average) amplitude.
This results in more stringent limits on the product of the neutrino
magnetic moment and magnetic field strength, $\mu_{\nu}B$.

\section{Neutrinos in random magnetic fields}
\label{sec:nurand}
Let us consider, for simplicity, the spin flavour precession of two
Majorana neutrinos in vacuum ~\cite{Schechter:1981hw}. The evolution
of the system is governed by the Schr\"odinger-like equation
\begin{equation}
i\:{\vec \nu} = {\cal H}\:\vec\nu,
\label{eq:main}
\end{equation}
where $\vec\nu^{T}=(\nu_{1},\bar\nu_{2})$, $\nu_i$ are neutrino mass states,
\begin{equation}
{\cal H} =\mu_{\nu}B_x\:
\sigma_1-\mu_{\nu}B_y\:\sigma_2-\delta\cdot\sigma_3 
\end{equation}
is the $(2\times2)$ Hamiltonian, $\sigma_j$ -- Pauli matrices,
$\mu_{\nu}$ -- the neutrino transition magnetic moment, $B_x$ and
$B_y$ are magnetic field components perpendicular to the neutrino
trajectory (along $z$ axis), and $\delta=\Delta m^2/4E$; $E$ and
$\Delta m^2$ are the neutrino energy and the squared mass difference,
respectively.

In a uniform magnetic field the conversion probability is
\begin{equation}
\label{eq:regular-probs}
P\left(\nu_{1}\to\bar\nu_{2};L\right)=
\frac{\mu^2_{\nu}B^2_{\perp}}
{\delta^2+\mu^2_{\nu}B^2_{\perp}}\sin^2
\left(\sqrt{\delta^2+\mu^2_{\nu}B^2_{\perp}}L\right),
\end{equation}
where $L$ is the neutrino path in the magnetic field
region. Hereafter we will not distinguish $z$ and $t$ assuming
$z=t$ for $c=1$.

Let us now assume that a probe neutrino crosses a region with a
small-scale random magnetic field with effective scale $L_0$.  The
neutrino trajectory is divided into about $N=L/L_0$ correlation cells.
For a given realization, the random magnetic field vector is assumed
to be uniform within each cell; the fields in adjacent cells are
uncorrelated and, moreover, within one cell different magnetic field
components, transversal to the neutrino trajectory, are also
independent random (Gaussian) variables with zero mean
value~\cite{Bykov:1998gv}.

In the uniform magnetic field case the evolution matrix
$U=\exp\{-i{\cal H}\cdot z\}$ is trivially found. For any
piece-constant magnetic field profile (set of field domains) it is
then just the product of corresponding unitary matrices,
\begin{equation}
\label{multipl}
U(L)=\prod^{N}_{j=1}U_j,
\end{equation}
where
\begin{equation}
U_j=\exp(-i{\cal
H}L_0)=\cos\omega_j-i(\vec\sigma\cdot\vec{n_j})\sin\omega_j 
\end{equation}
with
\begin{equation}
\begin{array}{cc}
\omega_j={\cal D}_j\cdot L_0, & {\cal D}_j = \sqrt{\delta^2+\mu^2B^2_{j\perp}},\\
\vec n_j=(\mu_{\nu}B_{jx}, -\mu_{\nu}B_{jy}, -\delta)\cdot{\cal
D}_j^{-1} & \vec n^2_j=1.
\end{array}
\end{equation}
The neutrino conversion probability after crossing the random magnetic
field region is therefore equal to the corresponding matrix element,
\begin{equation}
P(\nu_{1}\to\bar\nu_{2};L)=
\langle 0|U_1^*U_2^*...U_N^*\frac{1-\sigma_3}{2}U_N...U_2U_1|0\rangle, 
\end{equation}
where $|0\rangle^T=(1,0)$ is the initial neutrino state and
$(1-\sigma_3)/2={\rm diag}(0,1)$ is the projector on to $\bar\nu_{2}$
state. Because of the multiplicative nature of the evolution matrix,
Eq.(\ref{multipl}), we perform the averaging of the conversion
probability step by step. After commutation we obtain the following
inner matrix structure
\begin{equation}
U_N^*\frac{1-\sigma_3}{2}U_N=\frac12[1-\cos(2\omega_N)\cdot\sigma_3-
\sin(2\omega_N)
\cdot[\vec n_N\times\vec\sigma]_3-2\sin^2\omega_N\cdot
n_{N,3}(\vec\sigma\vec n_N)].
\label{eq:matrix-rel}
\end{equation}
Taking into account that averaging over random magnetic fields in the
$N$-th cell washes out all terms proportional to odd powers of
$B_x^{(N)}$ and $B_y^{(N)}$, we obtain
\begin{equation}
\left\langle U^*_N\frac{1-\sigma_3}{2}
U_N\right\rangle_{\mbox{av}}=\frac12\left[1-\left(1-2\left\langle
\left(1-n^2_{N,3}\right)\sin^2
\omega_N\right\rangle_{\mbox{av}}\right)\sigma_3\right],
\end{equation}
that is just the same diagonal matrix as the initial projection
operator modified only by a scalar factor in front of $\sigma_3$.

Therefore by induction and after some algebra we obtain
\begin{equation}
\label{eq:ave-probs}
\left\langle P(\nu_{1}\to\bar\nu_{2};L)\right\rangle_{\mbox{av}}
=\frac12-\frac12\prod^N_{j=1}\left(1-2P_j^{(c)}\right),
\end{equation}
where $P_j^{(c)}$ is the averaged conversion
(flavour-changing) probability in the j-th correlation cell given
by
\begin{equation}
P_j^{(c)}=\left\langle
\frac{\mu^2_\nu B^2_{j\perp}}{\delta^2+\mu_\nu^2 B^2_{j\perp}}
\sin^2\left(\sqrt{\delta^2+\mu^2_{\nu}B^2_{j\perp}}L_0\right)\right\rangle_{\mbox{av}}.
\end{equation}
When all conversion probabilities are small, i.e. when
$\mu^2_{\nu}\left\langle B^2_{\perp}\right\rangle\ll\delta^2$ (and this is the case for
solar neutrino oscillation parameters and realistic magnetic
fields, see below), Eq.(\ref{eq:ave-probs}) is greatly simplified,
\begin{equation}
\left\langle P(\nu_{1}\to\bar\nu_{2};L)\right\rangle_{\mbox{av}}
\approx\sum^{N}_{j=1}P_j^{(c)}= \sum^N_{j=1}\frac{\mu^2_{\nu}\left\langle
B^2_{j\perp}\right\rangle}{\delta^2}\sin^2\left(\delta\cdot L_0\right).
\end{equation}

The above results mean that because of the randomness of magnetic
fields the neutrino spin-flavour evolution looses coherence, that is
instead of dealing with wave functions it is necessary to consider
probabilities. Therefore for small conversion the resulting effect is
of cumulative nature and the probability is proportional to the number
of correlation cells of the random magnetic field.

Let's assume that all root-mean-square random field amplitudes in
different cells are equal to the strength of some constant regular
magnetic field. In this case we see that the above result is
proportional to the number of correlation cells traversed by neutrino,
$N = L/L_0$, thus leading to a sizable gain in neutrino conversion in
random field as compared with the case of a constant magnetic field of
the same amplitude. Indeed, in the case of regular field from
Eq.(\ref{eq:regular-probs}) we have
\begin{equation}
\label{regfield}
 P(\nu_{1}\to\bar\nu_{2};L) \approx
 \frac{\mu_{\nu}^2 B_{\perp}^2}{\delta^2}
\sin^2  \left(\delta L \right) + O\left(\left( \frac{\mu_{\nu}^2
B_{\perp}^2}{\delta^2}\right)^2 \right) \approx  \frac{\mu_{\nu}^2
B_{\perp}^2}{2\delta^2}~,
\end{equation}
that is similar to the case of neutrino passing only one cell of the
size $L$.

\section{Neutrinos in solar random magnetic fields}
\label{sec:solar}

The simplified approach given above can be taken over to the general
case of the neutrino spin flavour precession in solar random magnetic
fields~\cite{Miranda:2003yh}.
Within this generalized picture (LMA-MSW + SFP), after the MSW flavour
conversion occurred in the inner region of the Sun, $\bar{\nu}_{e}$'s
are produced due to the magnetic moment conversion $\nu_{\mu} \to
\bar{\nu}_{e}$ in the convective zone magnetic field. The two-flavour
Majorana neutrino evolution Hamiltonian in matter and magnetic field
is well--known to be
four--dimensional~\cite{Schechter:1981hw,akhmedov:1988uk}.  However
for solar convective zone random magnetic fields the full $4\times4$
evolution equation decouples into two $2\times2$ equations describing
LMA-MSW oscillations deep in the Sun and the following (approximate)
vacuum SFP conversions inside the solar convective
zone~\cite{Miranda:2003yh}.  This is explained by smallness of two
main parameters, $V/\delta \simeq 10^{-2}$ , where $V$ is the matter
potential within the convective zone, and
\begin{equation}
\kappa = \frac{\mu_\nu^2 b_\perp^2}{\delta^2} = 2.5 \times 10^{-5}
\left(\frac{\mu_\nu}{10^{-11}\mu_B}\right)^2 \left(\frac{b_{\perp
max}}{100 {\rm kG}}\right)^2 \left(\frac{7\times 10^{-5} {\rm
eV^2}}{\Delta m^2}\right)^2 \left(\frac{E}{10{\rm MeV}}\right)^2~.
\label{kappa}
\end{equation}
The first parameter tells us that the matter effects inside the
convective zone are negligible and can be safely neglected. The
smallness of the parameter $\kappa$ allows to use the perturbative
approach described in Section~\ref{sec:nurand}.

\section{Results and Discussion}
\label{sec:results}

From the above discussion one sees that spin flavor conversion is much
more efficient in producing solar anti-neutrinos for random magnetic
fields than for the case of regular fields. To confirm our conclusions
numerically we compute the limits on $\mu_{\nu}B$ both for regular and
for random magnetic fields. The results are plotted in
Fig.~\ref{fig:reg-rnd}.
\FIGURE[t]{\epsfig{file=rnd-reg.eps, width=.8\textwidth}%
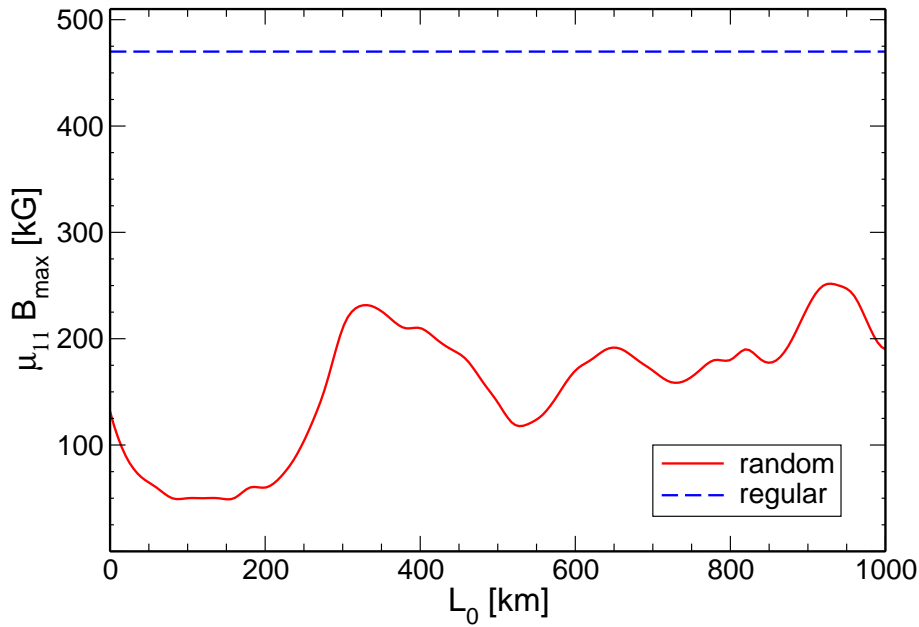
            \caption{Bounds on $\mu_{11} B_{max} $ for random
              magnetic field versus correlation scale $L_0$ (solid
              line). The horizontal dashed line indicates the bound on
              $\mu_{11} B_{max}$ for Kutvitsky-Solov'ev regular
              magnetic field. $\mu_{11}$ is magnetic moment in units
              of $10^{-11}$ Bohr magneton. Details are given in
              text.}\label{fig:reg-rnd}}
The full available set of neutrino data was taken into account along
with recent KamLAND bound on the electron anti-neutrino
flux~\cite{Eguchi:2003gg}. To make connection with previous results
the Kutvitsky-Solov'ev magnetic
field~\cite{Barranco:2002te,Kutvitskii} was taken as a reference
regular field as well as the root-mean-square random field shape. Here
the correlation scale $L_0$ was considered as an additional free
parameter.
For regular fields we obtain the constraint 
\[ 
\mu_\nu B_{max}<10^{-11}\mu_B \times 470~\mbox{kG at 90\% C.L.}
\] 
On the other hand for the random magnetic field case one finds, in the
most conservative case,
\[
\mu_\nu B_{max}<10^{-11}\mu_B \times 250~\mbox{kG at $L_0\sim 950$~km
(90\% C.L)}
\]
while for the most optimistic case 
\[
\mu_\nu B_{max}<10^{-11}\mu_B \times50~\mbox{kG at $L_0\sim100$~km
(90\% C.L).}  
\]
If we further specify the random magnetic field model to be of the
turbulent type one can eliminate a dependence upon the correlation
scale since neutrinos effectively feel only one scale with the space
period equal to the neutrino oscillation length~\cite{Miranda:2003yh}.

Taking into account that the present-day understanding of the solar
magnetic field evolution leads to small-scale convective zone random
magnetic fields comparable or even exceeding the large-scale ones, we
can conclude that the former indeed can play an important role in the
analysis of the solar neutrino data.

\acknowledgments

We thank the organizers for the warm and fruitful atmosphere during
the AHEP conference.  We thank V. B. Semikoz and D.  D.  Sokoloff for
useful discussions.  This work was supported by Spanish grant
BFM2002-00345, by European RTN network HPRN-CT-2000-00148, by European
Science Foundation network grant N.~86 and MECD grant SB2000-0464
(TIR). TIR and AIR were partially supported by the Presidium RAS
program ``Non-stationary phenomena in astronomy'' and CSIC-RAS
exchange program.  OGM was supported by CONACyT-Mexico and SNI.

\end{document}